# Towards a Theory of Modular Natives: Explaining Superscaling, China's Greatest Innovation Yet[1] [2]

By

Bent Flyvbjerg,[3] Alexander Budzier,[4] and Maria Christodoulou[5]

The University of Oxford

IT University of Copenhagen

---

[1] Division of work: Bent Flyvbjerg coined the term "modular native," developed the theory of modular natives and superscaling, tested it, and wrote the paper. Alexander Budzier contributed to data collection, statistical analyses, and commented on an earlier draft of the paper. Maria Christodoulou did the statistical analyses and commented on an earlier version.

[2] Acknowledgments: The authors wish to thank John Kwong for facilitating and participating in field visits to China to study modularity and superscaling at work, Dr. Jens Schmidt for comments on an earlier version of the paper, and the Villum Foundation for financial support.

[3] Professor Emeritus, University of Oxford's Saïd Business School; Villum Kann Rasmussen Professor and Chair, IT University of Copenhagen; Senior Research Fellow, St Anne's College, Oxford University, flyvbjerg@mac.com (corresponding author).

[4] Fellow in Management Practice, University of Oxford's Saïd Business School.

[5] Senior Statistical Consultant, Department of Statistics, University of Oxford.

**Highlights**

- We coin the term "modular native," defined as a basic building block that is born modular, e.g., a solar cell.
- We demonstrate that modular natives are key to success in building things, while non-natives spell trouble.
- China understands this better than any other geography, which explains its global success as superscaler of renewables, EVs, robots, and more.
- Modular natives and superscaling are major inventions in their own right, falsifying the notion that China cannot innovate.
- Modular natives are main drivers of prosperity and geopolitical power.
- China presently sets the benchmark for modular natives and superscaling, against which other nations must measure themselves.

**Abstract**

First, we present a new theory of "modular natives." A modular native is a basic building block that is born modular, e.g., a solar cell. The theory predicts that using modular natives in building things reduces complexity and improves predictability, resulting in better outcomes and faster scale-up. Second, we test the theory on the largest dataset of its kind. We find, at a high level of statistical significance, that modular natives operate under a fundamentally different risk regime than other project types, with finite and predictable risk, in contrast to non-natives that have infinite and unpredictable risk. The findings help explain why modularity is key to successful building while bespokeness often leads to failure. Third, we relate our findings to economic and geopolitical development, arguing that China understands modular natives and scale-up better than any other geography and that this is key to China's swiftly growing dominance in renewables, batteries, EVs, robots, etc. We argue that China's mastery of modularity and scale-up is a major innovation in its own right, among the greatest and most impactful in human history, falsifying the common notion that China cannot innovate. Business and government outside China ignore these findings at their peril. Finally, we spell out policy and practice implications and identify areas for further research.

**Introduction**

In November 2025, *The New York Times* reported a "surprising shift" in the expansion of renewable energy to fast-growing developing nations where most people live and where renewables may more effectively impact climate than anywhere else (Sengupta and Plumer 2025). Brazil, India, and Vietnam were rapidly growing solar and wind power. Nigeria was planning to build its first solar-panel manufacturing plant. Morocco was creating a battery hub to supply Europe with electricity. Nepal and Ethiopia were leapfrogging petrol-burning cars for electric ones. All driven by "the world's new renewable energy superpower: China," according to the *Times*.[6]

A London think tank similarly reported that in the first half of 2025, renewables overtook coal in global electricity production for the first time (Wiatros-Motyka and Rangelova 2025). This happened via record growth in solar and wind power, again lead by China. In just a few years, the country had scaled up production exponentially while driving down cost, also exponentially, making solar and wind the cheapest and fastest growing sources of energy in the world. In the process, China met several of their climate goals ahead of schedule, in a world where infrastructure is almost always delivered late and over budget (Flyvbjerg and Gardner 2023). No less important, China now dominated the market for renewables, with a global market share for the production of solar panels of over 80 percent, and over 60 percent for wind turbines, more remarkable for the fact that modern solar and wind power were both invented in America and Europe.

[6] At the same time, The Economist (2025a: 50) wrote that 2024 was "the greenest year ... in the BRI's history," the BRI being the Belt and Road Initiative, China's giant infrastructure-building scheme, to which most developing nations have signed up.

How did China pull off this extraordinary feat, at a scale and speed never seen before in human history? How did America and Europe lose such important industries to China so quickly? And why did something similar happen shortly afterwards for batteries, electric vehicles (EVs), drones, robots, and more?

Below, we explain China's success by a new theory called the *theory of modular natives*. According to the theory, modularity enables scale-up, especially a particular form of modularity, which we call *modular natives*. Modular natives drive an extreme form of scale-up, which we call *superscaling*. We argue that China understands modular natives and superscaling better than any other geography and that this is a main reason why China was able to dominate solar, wind, and other industries so quickly. Finally, we argue that China's mastery of modularity and superscaling is a *major innovation* in its own right, falsifying the common notion that China cannot innovate and therefore is no real competition to America and Europe in the economic and geopolitical world order.

First, we explain the theory of modular natives. Second, we describe the data and methodology used to test the theory. Third, we present our findings, and especially how modular natives facilitate building things, whether in manufacturing, services, or infrastructure. Finally, we discuss the implications of our findings, including for competition in the global world order between China, America, and Europe, economically and geopolitically.

**The Theory of Modular Natives**

We follow Baldwin and Clark's (2000: 63) classic definition of a module as "a unit whose structural elements are powerfully connected among themselves and relatively weakly connected to elements in other units."[7] A defining characteristic of modules is, "that they are independent of one another, constrained only by their adherence to a common set of design rules" (op. cit., p. 283).[8] Stated in mathematical-statistical terms, in a modular design variance manifests mainly within modules and less between them. Making something modular is therefore a way to control variance, that is, risk, by containing it inside modules.

We similarly define *degree of modularity* as the extent to which something can be made modular, with "the ideal of 'perfect' modularity" at one end of the scale and fully bespoke at the other (Baldwin and Clark's 2000: 89). Between these two endpoints, intermediate types exist, making modularity a matter of degree, not an either-or. The theory of modularization says that *the higher the degree of modularity you can achieve, the more effectively you will be able to control complexity, variance, risk, and outcomes*.[9]

We coin and define the term *modular native* to describe products, services, and infrastructure that incorporate Baldwin and Clark's "perfect modularity" by design, that is, from their conception. The term is inspired by the concept of a "digital native," which Prensky (2001a, b) defined as a person who was born digital in the sense that they were conceived and grew up in the digital age (Selwyn 2009). To illustrate, the shipping container is an archetypical modular native, with the individual container as the basic building block of the worldwide shipping industry. So is a Lego set, with the individual Lego brick as the basic building block. And so is a solar panel, with the solar cell as the basic building block, from which solar arrays, and solar farms can be built in modular fashion at increasing levels of so-called "scale-free scalability," meaning that each level can be scaled up further based on the same basic module and the same design rules by simple repetition: solar cell repeated = solar panel; solar panel repeated = solar array; solar array repeated = solar farm; solar farm repeated = solar industry, all based on repetition of the same basic building block, the solar cell (Flyvbjerg 2021: 9). We theorize that modular natives are particularly interesting as an object of study – more interesting than modularity as such – because they are a type of modularity that increasingly makes a difference economically and geopolitically.

---

[7] Here we focus on human-made modules. It is interesting to note, however, that modularity is increasingly recognized as a fundamental feature of nature that may have profound consequences for biology and evolution (Kurzban and Aktipis 2007, Wagner et al. 2007, West 2017, Zelditch and Goswami 2021).

[8] For variations on this definition, see Campagnolo and Camuffo (2010).

[9] Modularization may be seen as a particularly rigorous kind of standardization, which we define as the process of bringing technical standards to the development and deployment of technology (Tassey 2000, Shin et al. 2015).

The opposite of modularity is *bespokeness*, meaning something tailor-made for a specific context, typically as a one-of-a-kind, that is, unique (Flyvbjerg et al. 2025b). English handmade shoes are bespoke, made-to-measure uniquely for the feet that will wear them. So are most nuclear power plants and hydroelectric dams. Building something bespoke is slow and complex, involving lots of gauging, sizing, and variance, which translates into risk and unpredictability (Baldwin and Clark 2000).

The *theory of modular natives* hypothesizes that modularity is less risky than bespokeness in building things, specifically that modular natives are an effective way of controlling complexity, risk, and predictability and thus outcomes. The theory predicts that *the more modular natives are used in the building of things, the less risky and the more predictable those things will be, with better outcomes*, other things being equal. Below, we test the theory by empirically testing the following null hypothesis:

> *$H_0$: Modular natives perform no different from other artifacts in terms of risk and predictability.*

Such test has not been done before, just as the concept of modular native and the theory of modular natives are proposed here for the first time.

Finally, we spell out what our findings mean for policy and practice. Given the trillions of dollars spent yearly on building things, the economic and financial implications are massive as to whether this is done with more or less risk and predictability, i.e., more or less effectively.

**Data and Methodology**

In this section we present the data and methodology used to test the theory of modular natives. To our knowledge, no ideal dataset exists for testing whether modular natives are different from other artifacts in terms of risk and predictability in building things. However, in our research on project planning and management we have built a large database covering more than 400,000 data points measured on over 16,000 projects across 23 project types (Flyvbjerg et al. 2026). These data may serve as a useful first step in testing the thesis. They include the type of solar and wind power projects mentioned in the introduction. Projects are also a particularly pertinent perspective on the problematic at hand, because the design and delivery of projects account for a large and increasing share of the world economy to a degree where the process by which this is happening has acquired its own name, namely "projectification," resulting in what has been called the "project economy," estimated at USD 20 trillion in 2027, up from USD 12 trillion in 2017 (Nieto-Rodriguez 2021).[10]

---

[10] Projectification entails less emphasis on traditional, permanent organizational structures and processes and more on temporary activities defined as projects, with their focus on goals, timelines, resources, scope, and outcomes, and with reliance on ad hoc project-based teams and collaborations. Projects may be – and are – designed as modular or bespoke or intermediate types.

We collected data on three main types of project performance: Cost risk, schedule risk, and benefit risk. Here we focus on cost risk. The dataset has been used previously in, e.g., Flyvbjerg and Bester (2021) and Flyvbjerg et al. (2026), but never for testing theses about modularity. Cost risk and cost data are considered particularly important in this context because delivering projects within budget is a key criterion for project performance and project success (Baccarini, 1999).

Our data covered the 23 project types listed in Table 1, from housing to transportation, energy to water, information technology to defense, aerospace to mining. Cost risk could be systematically compared in a like-for-like fashion across 11,011 projects. Geographically, the projects are in 126 countries on six continents. Historically, most projects have been completed recently, in the 21st century. The average project size ranges from USD 35 million to 19.5 billion in 2025 prices. Average project duration varies from one to 14 years, measured from the final investment decision to going live. The data cover both government and business projects. The total value of the project portfolio is USD 4.97 trillion in 2025 prices. The sample is several times larger than the average sample size found in studies of project performance (Flyvbjerg et al. 2022).

We follow standard practice going back to Pickrell (1990) and measure cost risk by cost overrun, defined as actual divided by estimated cost in real terms, baselined in the estimate at the time of the final investment decision (FID).[11] A value of one represents on-budget delivery, above one over budget, and below one under budget. For example, the value of 1.52 shown in Table 1 for defense projects indicate an overrun of 52 percent for that project type, on average, in constant prices and measured from the FID until going live.

---

[11] Consistency of baselines across project types is crucial for the interpretation of data and results. Baselining at the FID measures whether decision makers were well informed or not when making the decision to build and avoids well-known biases related to underbidding when baselining at tendering. For details of the many pitfalls that exist for measuring cost overrun, including baselining, see Flyvbjerg et al. (2018).

To measure cost overrun on individual projects, we collected data on estimated and actual cost for each project. The data were collected from two main sources: (a) directly from owners of projects, based on their accounts, and (b) from public sources, e.g., auditor reports, official reports, and academic research. Only high-quality data, i.e., data based on valid and reliable written records, were included, which means we did not collect data from less reliable sources such as interviews, questionnaires, newspaper articles, and non-peer-reviewed conference papers to avoid the biases and errors associated with such sources (Flyvbjerg et al. 2018). For analysis, we included project types for which we had data for ten or more projects, 23 types in total.

We did conventional descriptive analysis of average and median risk based on all 11,011 projects and all 23 project types. For the analysis of tail risk, which turned out to be the more important type of risk, we focused on the upper (right) tail, i.e., projects with an overrun above 1, amounting to 5,596 projects, again for all 23 project types. We built on Clauset et al. (2009) to fit probability distributions to the data. We added visual inspection using quantile-quantile (QQ) plots of probability distributions for each project type. We supplemented this with Pareto and lognormal goodness-of-fit tests using the Anderson-Darling test, the Chi-square test, and the Approximate Kolmogorov-Smirnov test. Finally, we did a non-parametric bootstrap analysis of the tail parameter, alpha, before settling on the Pareto 1 as the best fit. For detail, we refer to Flyvbjerg et al. (2026).

We took measures to ensure robust findings, i.e., that results would not be artifacts of how we collected, classified, or analyzed data.[12] First, we ensured that only high-quality data were included in the study, as mentioned. Second, we made sure to collect data from different types of sources and tested whether our findings were robust across those, which was the case. This test was particularly important because differences in variance between different data sources may spuriously generate fat tails, which would then be an artifact of data collection instead of a characteristic of the data. We established that the fat tails we found were not artificial in this manner. Third, we tested for robustness of results across project size, project duration, and year of going live, again to ensure that the fat tails identified were not generated as artifacts of mixing data that should not have been mixed, i.e., data from different project populations. Fourth, we tested different ways of categorizing the data according to project type and found that cost overrun behavior is not an artifact of how projects are categorized. Finally, whenever different statistical methods existed for testing our hypotheses, we used each of them to test whether our results were robust across methods, which was the case. In sum, the robustness tests provide strong evidence that the observed patterns in cost overrun reflect real properties of the underlying data and are unlikely to be artifacts of data collection, categorization, or analytical choices.

[12] The robustness tests were primarily conducted at the level of the 23 project types and are therefore not specific to the binary aggregation into modular natives versus other project types used below. The tests are still relevant and are included because they establish that the cost overrun distributions themselves are not artifacts of data source, project size, project duration, time period, or project taxonomy (for details of the robustness tests, see Flyvbjerg et al. 2026). The aggregation into two groups does not introduce statistical inconsistencies and, if anything, represents a conservative analytical choice, because aggregation reduces variance and statistical power rather than inflating differences.

Finally, we note that the data may have a conservative selection bias – i.e., projects in the project population may perform worse than projects in the sample. This is because, as mentioned, we collected data only on projects for which high-quality written records were available. But, organizations that keep high-quality records and are willing to share those for research purposes are generally considered more mature in their management practices than average organizations, producing better outcomes. This bias should be kept in mind when interpreting the findings below.

Further details about data collection, statistical analyses, and robustness tests may be found in Flyvbjerg et al. (2022, 2026).

**Findings: Less Complexity and Risk, Better Outcomes at Scale**

Table 1 shows the mean and the tail (defined as the percentage of projects with a cost overrun larger than 1.5, that is, 50 percent) for each of the 23 project types in the dataset. Projects are ranked by mean cost overrun, from smallest to largest. The larger the mean and the larger the tail, the poorer the quality of project performance in terms of cost; and the lower the better, with zero overrun (1.0 measured as a ratio) indicating that a project was delivered on budget.

*Table 1: Project types ranked by mean cost overrun, in real terms measured from the final investment decision, N = 11,011.*

| Project type | N | Mean cost overrun | % of projects with overrun > 1.5 |
|---|---|---|---|
| Solar power | 41 | 1.01 | 2.44 |
| Electricity transmission | 50 | 1.08 | 4.00 |
| Wind power | 84 | 1.12 | 7.14 |
| Pipelines | 437 | 1.14 | 9.61 |
| Roads | 2086 | 1.15 | 9.97 |
| Fossil thermal power | 136 | 1.18 | 15.44 |
| Water | 265 | 1.20 | 13.21 |
| Bridges | 61 | 1.26 | 21.31 |
| Mining | 886 | 1.27 | 16.93 |
| Oil and gas | 116 | 1.31 | 19.83 |
| Nuclear decom. | 16 | 1.34 | 25.00 |
| Tunnels | 69 | 1.36 | 27.54 |
| Rail | 372 | 1.39 | 27.15 |
| Rail stations | 71 | 1.43 | 23.94 |
| Defense | 115 | 1.52 | 20.00 |
| Buildings | 186 | 1.58 | 28.49 |
| Aerospace | 97 | 1.60 | 42.27 |
| Dams (other) | 21 | 1.71 | 33.33 |
| IT | 5360 | 1.73 | 18.26 |
| Hydroelectric dams | 300 | 1.75 | 37.33 |
| Nuclear power | 196 | 2.20 | 54.59 |
| Olympics | 21 | 2.57 | 76.19 |
| Nuclear storage | 25 | 3.33 | 52.00 |

We see large variation in overrun, and thus in the quality of performance, with the biggest overrun being no less than 233 times larger than the smallest. Nuclear storage projects – i.e., projects that stow nuclear waste from nuclear power plants or nuclear weapons – have the highest mean cost overrun at 233 percent (3.33 as a ratio), i.e., on average they overrun by more than a factor three. Being an average, this means that many individual nuclear storage projects have overruns larger than this. In contrast, solar power projects, i.e., projects that build solar farms, have the lowest mean overrun at one percent (1.01 as a ratio), indicating that many such projects were in fact delivered on or under budget. Similarly, the fattest tail in Table 1, for the Olympic Games (76.19 percent), is 31 times fatter than the least fat tail, again for solar (2.44 percent).[13] In all cases, cost overrun is measured in real terms, that is, without inflation, baselined in the final investment decision. Consequently, the numbers shown are conservative. If inflation were included and if the baseline for measuring overrun was a draft business case or an early feasibility study, the numbers would be significantly higher.

Visual inspection of Table 1 reveals an interesting phenomenon: The project types at the top of the table, with good performance (low overrun), are modular, while those at the bottom, with bad performance (high overrun), are generally bespoke. For statistical test, we would have liked to measure the degree of modularity of each project type and then correlate this measure with project performance. However, although the concepts of module and modularity are well defined and well understood, as described above, unfortunately no one has yet devised a rigorous and generally accepted methodology for accurately measuring an artifact's degree of modularity, to the best our knowledge. This is an important area for further research (Cabigiosu and Camuffo 2016).

We therefore attempted a workaround. Instead of measuring the degree of modularity for each project type on a ratio scale, which would be ideal, we decided to measure it on a nominal scale, with two categories: (a) modular native and (b) other. We argue that the top four project types in Table 1 (solar power, electricity transmission, wind power, and pipelines) are as close to Baldwin and Clark's "perfect modularity" as projects get in practice, i.e., they are modular natives and we classify them as such.[14] For the other 19 project types, many have elements of modularity – roads and thermal power plants, for instance – but they also contain significant elements of bespokeness, sometimes very much so, like the Olympic Games and nuclear storage. Because we have no valid and reliable way of measuring the varying degrees of modularity across the 19 project types, we grouped them together in the nominal category "other." We then tested the performance of modular natives against other projects, measured by cost overrun.[15] Specifically, we tested the null hypothesis mentioned above, "modular natives perform no different from other artifacts in terms of risk and predictability."

---

[13] We analyze the tails in more detail below.

[14] For solar projects, the basic building block is the solar cell, as already argued. For electricity transmission, it is transmission lines, transformers, and substations. For wind power, it is wind turbines, each made up of a foundation, a tower, a nacelle, and blades. Finally, for pipelines the basic building blocks are pipes, fittings, valves, and pumps.

Figure 1 shows the distribution of cost overrun for modular natives versus other project types (non-natives). In mathematical-statistical terms, the diagram shows the cumulative distribution function (CDF) for each of the two categories of project, plotted in a log-log diagram.

Eyeballing Figure 1, we see that overrun is substantially smaller for modular natives than for other project types. Specifically, mean overrun for modular natives is 12 percent, compared with 54 percent for non-natives, i.e., 4.5 times higher for non-natives than for natives, measured by the mean.

If we focus on projects in the upper tail (extreme overrun), here defined as projects with a cost overrun larger than 50 percent, then 8.3 percent of modular natives fall into this category, while 18.7 percent of other project types do. In other words, the risk of extreme overrun is 2.3 times larger for other project types than it is for modular natives. The difference between the two tails with overrun larger than 50 percent is statistically significant (Kruskal-Wallis test, statistic = 4.55, $p < 0.001$).

---

[15] We could also have chosen to test modular natives against the most bespoke projects in the sample, e.g., nuclear storage, the Olympics, and nuclear power projects, to study the difference in performance between modular natives and bespoke projects. However, in our judgment this would open the analyses to critique for arbitrarily including or excluding parts of the sample, which might impact results. We avoid this by including the whole sample, with every data point, in our analyses.

Finally, to understand the upper tails with more nuance, we fitted theoretical probability distributions to the data, using the Anderson-Darling test, the Chi-square test, and the Approximate Kolmogorov-Smirnov test, as described in detail in Flyvbjerg et al. (2026). We found the best fit for the Pareto 1 distribution and measured the tail parameter, alpha, for each tail. For modular natives we found an alpha value of 2.63, indicating finite mean and variance. Thus the law of large numbers converges for modular natives resulting in finite, predictable, and manageable risk. For other project types, we found an alpha value of 0.97, indicating infinite mean and variance. Here the law of large numbers does not converge resulting in infinite, unpredictable, and difficult-to-manage risk for such projects.[16] The two alpha-values are statistically significantly different (Mood's median test, chi-squared = 8.0285, $p = 0.0046$). Figure 1 and the established alpha-values point to the precise mechanism that makes modular natives work so well in building things, namely *significant tail risk reduction*. Specifically, we see that modular natives seem to shift risk toward more manageable tail behavior.

Table 2 summarizes the three findings. We see that modularity matters to performance. Not only do modular natives perform significantly better than other project types in terms of cost overrun. The estimates indicate that performance is fundamentally different for the two types of project. In mathematical-statistical terms, for modular natives the data indicate that the *law of regression to the mean* applies, allowing conventional statistical analyses and risk management. Risk is *mild*, in Mandelbrot's (1997) terms. For other project types, in contrast, the data suggest that the *law of regression to the tail* applies, with conditions that severely limit the effectiveness of conventional statistics and risk management (Flyvbjerg 2020). Here risk is *wild*, again in Mandelbrot's words.

---

[16] For $\alpha \leq 1$, the mean is infinite, designating infinite risk. Conventional statistics that assume the existence of the mean will be ineffective in this situation, as will conventional prediction, by which we understand forecasting based on the mean. The law of large numbers is too slow to converge for the mean. This is the most extreme risk that exists for Pareto 1 distributions, called "regression to the tail" by Flyvbjerg (2020). For $1 < \alpha \leq 2$, variance is infinite, which again means infinite and unpredictable risk. The law of large numbers will converge to the mean, in theory, but often too slowly for practical purposes. Finally, for $\alpha > 2$, mean and variance are both finite and the law of large numbers begins to converge resulting in regression to the mean.

It is difficult to exaggerate the importance of this finding because it uncovers the fact that modular natives and other project types operate under two fundamentally different tail-risk regimes: one of finite variance and thus predictability (natives), the other of infinite variance and therefore unpredictability (non-natives). At the level of root causes, this is why modularity is key to building successfully while bespokeness often leads to failure.

These findings are based on measurements on a nominal scale, as mentioned. Ideally, we would have preferred measurements on a ratio or interval scale, which would have allowed more granular quantification and analysis. This is not possible at present, as said. If it had been, we theorize that higher degrees of measured modularity would correlate positively with better project performance. For now, however, this is a hypothesis for future test, when measurement of degrees of modularity hopefully becomes possible.

In sum, our findings support the theory of modular natives, specifically that (a) projects built from modular natives seem less risky and more predictable than projects that are not, and (b) modular natives therefore seem an effective way of controlling risk and outcomes in the process of developing and building projects.

*Figure 1: Distribution of cost overruns for modular natives v. other project types. Overrun is smaller and less fat tailed for modular natives than for other project types. N = 11,011.*[17]

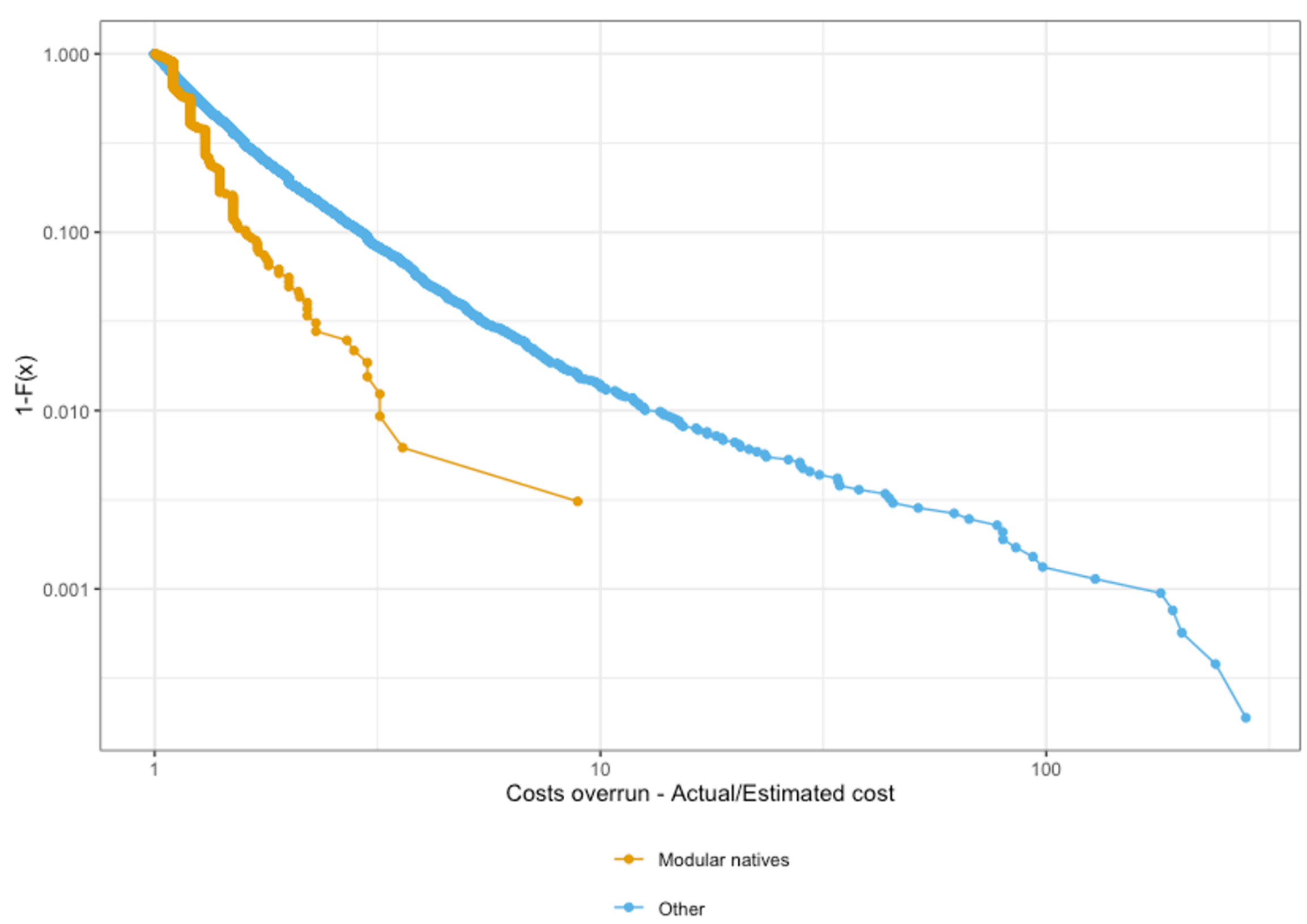

*Table 2: Cost overrun for modular natives v. other project types, measured by (a) mean, (b) percentage of projects with overrun above 50 percent (1.5), and (c) tail parameter, alpha, for the data shown in Figure 1 fitted with the Pareto I probability distribution in the upper tail, N = 11,011.*

| **Project type** | **N** | **(a) Mean** | **(b) % projects > 1.5** | **(c) Alpha** | **Statistical law** |
|---|---|---|---|---|---|
| Modular natives | 612 | 1.12 | 8.33 | 2.63 | Regression to the mean |
| Other | 10399 | 1.54 | 18.67 | 0.97 | Regression to the tail |

[17] In mathematical-statistical terms, the diagram shows the cumulative distribution function (CDF) plotted in a log-log diagram for modular natives and other project types, respectively.

**Discussion: A Wake-Up Call from China**

Here we discuss macro and micro implications of the theory of modular natives. First, at the macro level, the rise of China has conventionally been explained in terms of division of labor. America and Europe are seen as geographies of inventors, epitomized by Silicon Valley. China is seen as a geography of producers, typified by Shenzhen. Inventions flow from America and Europe, products from China. A famous Apple slogan – "Designed by Apple in California. Assembled in China" – neatly captures the logic. According to this, the West has nothing to fear from China, because "China can't innovate," as Abrami et al. (2015) famously claimed in *Harvard Business Review*, since China is mainly a producer. More recently, and even more bluntly, Arkansas senator Tom Cotton said on World IP day 2025, "China doesn't innovate – it steals."[18]

From the perspective of the theory of modular natives this logic is simplistic. It is also perilous for the West, economically and geopolitically. While Americans and Europeans have been repeating, like a mantra, that China cannot innovate, China invented the largest and fastest production and construction machine in history, right under their noses. China did this by understanding the type of data shown in Figure 1 and Table 2 better than any other nation and constantly driving production and construction towards higher degrees of modularity, with modular natives as the ultimate goal, because they best reap the benefits of speed and scale. We argue that this machine, and the resulting scale-up of solar, wind, batteries, etc. described above, is *a major invention in its own right, among the greatest and most impactful*

---

[18] Tweet on X by Tom Cotton, April 25, 2025, https://x.com/SenTomCotton/status/1915800089362079783.

*in human history*. We name it *superscaling*[19] and define it as the ability to exponentially grow products, services, and infrastructure in numbers and at speeds previously unseen, generating learning curves that quickly drive down costs. Modular natives and their learning curves are the building blocks of superscaling.

Learning curves are also known as "Wright's law," named after American aeronautical engineer Theodore Paul Wright, stating that the cost of a technology falls consistently as the cumulative production of that technology increases.[20] In the 1930s, as aircraft production ramped up, Wright observed that every time production doubled, the needed labor time for an additional aircraft fell by 20 percent, also called the *learning rate* (Wright 1936). Similar results have been found for other industries, with learning rates typically between five and 30 percent, depending on technology. E.g., for solar panels the learning rate is 20 percent, like for Wright's aircraft, resulting in a drop in cost of 90 percent over the past decade, with production doubling approximately every three years (Ritchie 2024). For onshore wind, the similar drop has been 70 percent, and for batteries, over 90 percent.[21] We see that superscaling is Wright's law on steroids, with superscaling making for faster cycles of learning, which make for lower costs.

---

[19] Similar terms found in the literature are "hyperscaling" and "blitzscaling" (Flyvbjerg 2021). Superscaling is not hyperscaling, however, which is the scale-up of computing systems, including server farms and AI, to handle vast workloads (Salahuddin et al. 2018). It is also not blitzscaling, which is the scale-up of organizations, typically tech companies (Hoffman and Yeh 2018). Superscaling is a more recent and more general engineering concept, cutting across ever more technologies and industries.

[20] Mathematically, Wright's law takes the form of a power function.

[21] Falling unit costs are sometimes also called "positive learning curves," which are contrasted with "negative learning curves," i.e., higher unit costs with larger production. Nuclear power, for instance, which is highly

We maintain that superscaling is one of the most transformative economic and geopolitical forces in the world today. Nevertheless, it is generally overlooked and not well understood, especially outside China. Invention and start-up have tended to get most of the attention here, by policy makers and scholars alike. That's a big mistake, because invention and start-up without scale-up is immaterial in today's world. Scale-up is ultimately what matters, and, especially, superscaling.[22]

Philanthropist John Arnold (2024) hit the nail on the head when he observed about this, "America has the ability to invent. China has the ability to build. The first country that can figure out how to do both will be the superpower." Arnold here spells out what may be considered the formula for present-day global power politics:

---

bespoke, has generally had negative learning curves that have hampered the spread of this technology, even in China that has tried harder and builds more nuclear power plants than any other country. Consequently, growth in new installed capacity for nuclear power is a fraction of what it is for solar and wind, both globally and in China, and the cost per megawatt is significantly higher (Flyvbjerg and Gardner 2023: 175-77).

[22] An Apple executive expressed the ultimate importance of scale-up in the following hands-on manner: "You can talk about touchscreen glass and the end of the BlackBerry keyboards all you want, but what really matters is having the genius to figure out how to produce a million of them each day" (McGee 2025: 316). Apple, in collaboration with Chinese contractors, had that "genius." Ironically, more than any other company, it is Apple – an American company – that has taught China how to superscale. But Apple could not have done this without close collaboration with giant contractors like Foxconn and Luxshare, on the ground in China, and without the low cost, long working hours, and government control of Chinese labor, not to speak of extensive subsidies. China may be the only economy in the world where Apple could have become the superscaler it is today, and where superscaling could have taken off at the speed and to the extent it now has across a wide variety of industries. McGee (2025: 385) calls Apple's role in Chinese superscaling "a major geopolitical dilemma," for both the company and the United States.

Invention + scale-up = economic and geopolitical power,

whereas invention minus scale-up equals irrelevance. However, Arnold is too pessimistic about Chinese invention, in our analysis. In fact, China has rapidly become a better inventor. It is fast hacking Arnold's formula to become the first global superpower to get both invention and scale-up right. Wong-Leung et al. (2025) reports that out of 74 technologies systematically tracked, based on high-impact research, "China now leads in 66 ... with the United States leading in the remaining eight." In consequence a reverse tech transfer is emerging, already visible in, e.g., battery storage, EVs, offshore wind, and solar power. "Invented in China, assembled in the U.S." is fast becoming as relevant as the old Apple slogan mentioned above. Tech transfer is rapidly turning into a two-way street.

Moreover, China's way to scale up – or build, to use Arnold's term – is a major invention in its own right, as already argued. It is a *meta-technology* that China presently wields more effectively than anyone else. But this only makes Arnold's wake-up call more important: If other countries want to maintain their position in the world, including their living standards, they must master modular natives and superscaling as well as or better than China, and currently they don't.

The theory of modular natives helps explain why and how superscaling works, namely by using modular natives as the basic units from which big things get designed and built. Modular natives entail significant risk reduction in the building process, from fat-tailed to less extreme risk, as illustrated by Figure 1. Less risk means less friction, allowing repetition and learning to a degree and at speeds not seen before. The theory also helps explain China's uniquely powerful position in the world today where, for the first time, the country is seriously challenging centuries of dominance by America and Europe. Finally, the theory makes clear that in meeting this challenge the West cannot rest on its supposed laurels in innovation. In sum, at the macro level the theory of modular natives sets out the winning formula for present-day markets and geopolitics: *superscale or fade*.

Second, at the micro level, which covers individual businesses, we find similar challenges. Tesla is possibly the company in the West that best understands and practices modularization and superscaling, both in the construction of its factories and the manufacturing of its products (Flyvbjerg 2021). Nevertheless, Tesla seems to initially have underestimated competition from China. In 2011, when Tesla was the undisputed leader in the EV market, Elon Musk dismissed and derided BYD as a competitor, scoffing "Have you seen their car?", according to *Fortune Magazine* (Mollman 2023). In 2023, with BYD selling 1.6 million EVs against Tesla's 1.8 million, Musk admitted that BYD was now "highly competitive" (op. cit.). Two years later, BYD was outcompeting Tesla, China dominated global EV sales, and Tesla seemed busy pivoting into robotaxis, AI, and robots (Gaudiaut 2025).[23] Like EVs, China dominates solar and wind power, as we saw above, and the story repeats for batteries, electricity transmission, drones, robots, satellite launches, with more to come. In sum, China dominates ever more business areas and world trade, producing the largest trade surplus ever reported by a country, at US dollars 1.2 trillion in 2025 (Bradsher 2026).

Again the theory of modular natives helps explain why. Chinese businesses, subsidized and otherwise supported by government, have been able to go further faster with modularization and superscaling than their competitors.[24] Yes, Tesla was an EV industry pioneer and leader. But then BYD (and others) came along, with access to a larger workforce with the relevant skills and a more supportive government, allowing superscaling to a degree that even Tesla could not keep up with.

---

[23] According to a recent report in The Economist (2025b: 55), there are "reasons to believe China may win the race to build a robotaxi industry at scale." First, in 2025, fifty Chinese cities allowed the testing of robotaxis, compared with half that in America. Second, China's urban population – the core market for robotaxis – is three times larger than America's. Third, the cost of a Chinese robotaxi is estimated at less than half of its American counterpart. Finally, state backing is strong in China.

The policy implications are clear. If western businesses want to compete with China they need to (a) be as good as Chinese businesses at modularization and superscaling or (b) find areas where superscaling is not yet possible, which are becoming ever fewer or (c) find areas of innovation that China does not yet master, as seems to be the strategy with the Tesla pivots mentioned above.[25]

We strongly advice business and government to acquaint themselves with theories of modularization, and especially the theory of modular natives as laid out and tested above. This would help to build better products, services, and infrastructure that can compete internationally. Modularity and superscaling are not either-or propositions but matters of degree, well suited for incremental improvements (Flyvbjerg 2021). The sooner business and government get started, the more they do, and on the more fronts, the less market share and geopolitical power they will have to cede, especially to China which has recently gained momentum in terms of superscaling and modular natives, leaving the rest of the world behind. Many things could go wrong in China, of course, both politically (e.g., social unrest, like under Covid) and economically (e.g., collapsing markets, like property recently, or debt spinning out of control). But for China's competitors to rely on this would be unwise.

---

[24] Although we do not know of reliable data on this, it is interesting to note that modular natives seem to be more generously subsidized in China than other technologies, probably because the government sees modular natives as more effective in achieving national economic and geopolitical goals.

[25] We recently asked the chairman of one of the ten largest chaebols in South Korea in which areas Korean business could still compete with China. He mentioned three areas in particular: High-end computer chips, pharmaceuticals, and aerospace.

We further recommend that executives, government officials, and scholars visit China to see modular natives and superscaling for themselves, in action. In our experience, people who have not done so don't fully understand these phenomena (including ourselves before we went). The scale and speed are simply too mind-boggling to be imagined. They must be experienced first-hand, inside factories and on the ground. Seeing is believing.

Finally, government, business, and NGOs will be unable to solve the problems of climate and decarbonization without massive and fast scale-up. China saw this before other nations and made a point of becoming world leader in renewables. But China did not stop there. It is dominating industry after industry: electric vehicles, batteries, solar power, wind power, electricity transmission, on-line shopping, on-line finance, high-speed rail, fabricated construction, and more. The United States has done well with invention and start-up but has lagged for scale-up. Europe is lagging in all three: invention, start-up, and scale-up.

Wang (2025) and Klein and Thompson (2025) sound wake-up calls for nations that don't build things anymore, including the United States and Europe. Such countries stand to lose out economically and geopolitically. In the case of major conflicts and war this would be particularly pronounced, because history shows that those who build best and fastest tend to win wars, or so these authors argue. We agree and offer the theory of modular natives to better understand (a) why building is currently particularly urgent for anyone competing in the global economy, in business or government, and (b) how to build better based on the principle of modular nativity.

**Further Research**

We identify four areas for further research. First, and most importantly, our test of the theory of modular natives should be replicated based on more and better data collected by other researchers. Our findings are clear, at a high level of statistical significance, to be sure. They must therefore be taken seriously until such a time new research would prove otherwise. However, one should not fully accept a theory based on just one dataset and one study. This is especially the case when the policy and practice implications of the theory are as important and far-reaching as they are for the theory of modular natives.

Second, more research is needed on how to measure modularity. Without solid methods for such measurement scholarly progress will suffer. Although the concepts of module and modularity are well defined and understood, unfortunately no one seems to have devised a rigorous and generally accepted methodology for accurately measuring an artifact's degree of modularity. We worked around this problem by using a nominal scale for measurement (natives v. non-natives). This is far from ideal, however. As a minimum, we would want to measure degrees of modularity on an ordinal scale, to obtain more detailed quantitative measurements, or, better yet, on an interval or ratio scale. Figuring out how to do this is another important area for further research. If a general methodology proves out of reach for now, then figuring out other workarounds for measuring modularity is also an area for further research.

Third, we need more and better data for further study. This is especially the case for the project types that currently have small samples (say, $N < 100$), including modular natives below this level (solar, transmission, wind), to help increase the quality of statistical analyses. But we also need data for additional project types, including for more modular natives (e.g., batteries, battery storage, and data centers), to test the robustness of results with more types. Finally, we need data from more geographies, to enable more granular analyses. Given the prominent place of China in the discussion above, more data from China would be particularly desirable.[26] We know from experience, however, that Chinese data are difficult to come by, especially data that are highly valid and reliable, as they must be for scholarly work (Ansar et al. 2016). Trying to crack the problem of obtaining more Chinese data is therefore also an area for further research.

Fourth, if we were to arrive at data and methods that would allow measuring degrees of modularity, further research would be obvious that correlate degrees of modularity with Wright's learning rates with a view to better understand how modularization may most effectively contribute to positive learning curves, and how learning may be accelerated, which is needed for dealing with pressing problems in climate, health, and more. Further research would also be useful for understanding which types of innovation best support modularization, and vice versa. Lastly, it would be useful to understand how these phenomena vary across geographic realms for specific technologies, in a more nuanced manner than China v. the rest of the world, to facilitate learning from those geographies that are particularly good at modularizing and superscaling specific technologies, just like we are beginning to learn from China today.

[26] The dataset for the present study includes 141 Chinese projects, out of the total sample of 11,011 projects.